# High-resolution thermal expansion of MgB$_2$

R. Lortz**, C. Meingast*, D. Ernst*, B. Renker*,
D. D. Lawrie***, and J.P. Franck***

*Forschungszentrum Karlsruhe, Institut für Festkörperphysik,
P.O.B.3640, D-76021 Karlsruhe, Germany
**Fakultät für Physik, Universität Karlsruhe, 76128 Karlsruhe, Germany
***Dep. of Physics, University of Alberta, Edmonton, Alberta, Canada T6G2J

*The thermal expansion of polycrystalline MgB$_2$ from 5-300 K is studied using high-resolution capacitance dilatometry. The thermal expansivity exhibits a small jump of - 5.8×10$^{-8}$ K$^{-1}$ at T$_c$ (in accord with expectations from the Ehrenfest relationship and published specific heat and pressure data) and a negative peak-like feature close to 5 K. No indications of any structural instabilities are observed.*

High-resolution thermal expansion measurements of MgB$_2$ are of interest for several reasons. First, the classical superconductors with the highest T$_c$s (A15 compounds) are notoriously close to a structural lattice instability[1] and it is therefore of interest if this is also the case for MgB$_2$. Recent neutron diffraction data of MgB$_2$ show some evidence for anomalous behavior of the lattice parameters near T$_c$, which were however at the resolution limit of that technique[2]. A small anomaly in the thermal expansivity of MgB$_2$ is expected at T$_c$ (Ehrenfest relationship) due to the negative pressure dependence of T$_c$[3], but this has not been observed so far. Second, recent specific heat measurements have shown that MgB$_2$ is not a simple BCS superconductor[4,5]; rather it appears that a two-gap model is necessary to describe the specific heat data[6]. Thermal expansivity offers another possibility to study this anomalous behavior because expansivity is thermodynamically closely tied to the specific heat.

Here we study the thermal expansion of MgB$_2$ samples using high-resolution capacitance dilatometry. Two different polycrystalline MgB$_2$ samples were studied, which gave very comparable results. We only present the data of the larger sample (L=8 mm) which are of higher quality. This sample was prepared by mixing Mg and B$^{11}$ powders, which were then pressed into a pellet and sintered in a quartz tube for 3 hours at 750 $^o$ C. The T$_c$ of this sample, determined using the dilatometer, was 37.8 K and approximately 1 K wide.

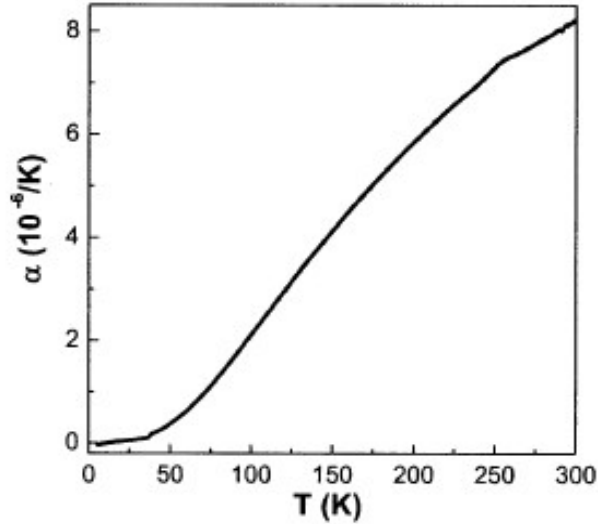

Fig. 1. Linear thermal expansivity of MgB$_2$ versus temperature.

In Fig. 1 we show the linear thermal expansivity $\alpha=1/L\times dL/dT$ from 5-300 K. The value of $\alpha$ at 300 K agrees well with the volume expansivity determined from neutron diffraction[2], implying that the grains in this polycrystalline sample are fairly randomly orientated. The expansivity on this scale appears nearly featureless, indicating that MgB$_2$ has no structural instability in the measured temperature range. We see no evidence for the quite large anomalies observed in the lattice parameters near T$_c$ (which would be very evident on the scale of Fig.1) observed in neutron diffraction[2].

The high resolution of our capacitance dilatometer however allows us to study the detailed response of the volume to superconducting order, and this is best demonstrated in a $\alpha/T$ versus T plot shown in Fig. 2. Here a clear jump $\Delta\alpha=\alpha_N-\alpha_S$=-5.8×10$^{-8}$ K$^{-1}$ is observed at T$_c$. Using the Ehrenfest relationship

$$\Delta\alpha = \frac{1}{3}\frac{dT_c}{dp}\frac{\Delta C_p}{T_c}$$

and values of the pressure dependence of T$_c$ (-1.11 K/GPa [3]) and specific heat jump at T$_c$ (2.1 mJ/moleK$^2$ [4] or 3.4 mJ/moleK$^2$ [5]), the thermodynamically expected jump in $\alpha$ at T$_c$ is calculated to be either -4.4 ×10$^{-8}$ K$^{-1}$ or $-7.2\times10^{-8}K^{-1}$, depending on which specific heat anomaly is used. Our experimentally determined value lies just between these two values, implying that the Ehrenfest relationship is obeyed quite well. This is another indication that no structural instability is involved in $MgB_2$, because in A15 and Chevrel superconductors, where structural transitions are known to play a crucial role, the Ehrenfest relationship is grossly violated[7,8].

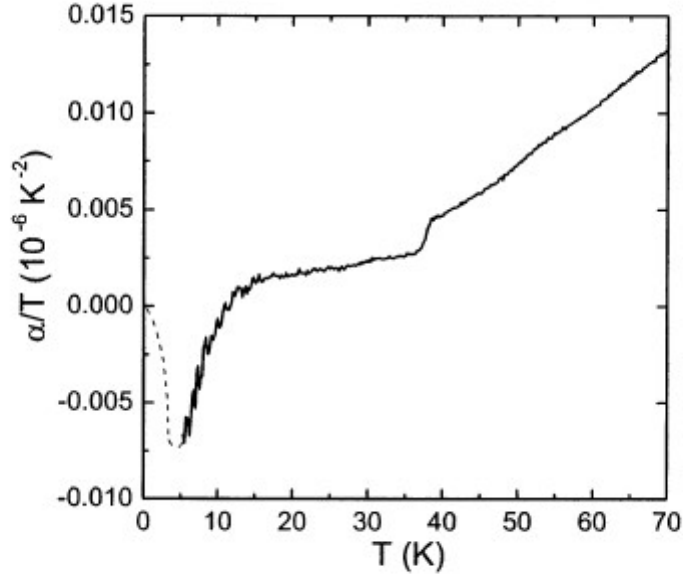

Fig. 2. $\alpha/T$ versus T of $MgB_2$ showing the anomaly at $T_c$. Below approximately 15 K, $\alpha/T$ exhibits a downturn and becomes negative to the lowest temperature of our measurement (see text). The dotted line indicates the expected behavior for a superconductor below 5K.

At lower temperatures, $\alpha/T$ exhibits a dramatic downturn, which continues to our lowest measured temperature of around 5 K. This feature occurs in the temperature range where an anomaly in the specific heat[4-6], which has been interpreted in terms of a phenomenological two-gap model[6], has been observed. The anomaly appears more prominent in the expansivity than in the specific heat probably because of a large Grüneisen parameter associated with the anomaly. Our data are, however, not fully consistent with the simple two-gap model, as explained below. $\alpha/T$ (just as $C_p/T$) of a superconductor is expected to go to zero as T approaches zero, and our data in Fig. 2 thus suggests that this low-temperature anomaly has a peak-like shape (dotted line in Fig. 2) rather than the step-like shape used in the specific heat model[6]. This possibly suggests that the two-gap model of independent gaps may be too simple to fully describe the thermodynamics of $MgB_2$. Expansivity measurements to lower temperatures and in magnetic fields are needed to further clarify this behavior.

In conclusion, a high-resolution thermal expansion study of $MgB_2$ show that: 1) The overall thermal expansivity is quite normal, i.e. there are no signs of any structural instabilities or phase transitions. 2) There is a small expansivity anomaly at $T_c$, as expected from the thermodynamic Ehrenfest relationship. 3) The thermal expansivity (just as the specific heat) exhibits an anomaly near $T_c/4$, which however does not appear to be correctly accounted for by the phenomenological two-gap model of Bouquet et al.[6].